\documentclass[%
 reprint,
 amsmath,amssymb,
 aps,
 prf,
 onecolumn,
 longbibliography,
]{revtex4-2}
\usepackage{graphicx}% Include figure files
\usepackage{dcolumn}% Align table columns on decimal point
\usepackage{bm}% bold math
\usepackage{gensymb}
\usepackage{textgreek}
\usepackage{xcolor}
\usepackage{hyperref}
\newcommand{\Vz}{V_{0}}
\newcommand{\muf}{\mu_{f}}
\newcommand{\Camuf}{Ca_{{f}}}
\newcommand{\betamax}{\beta_{max}}

\begin{document}

\preprint{APS/123-QED}
\title{Droplet impact on surfaces with asymmetric microscopic features}
% Force line breaks with \\

\thanks{A footnote to the article title}%

\author{Susumu Yada$^1$}
\email[corresponding author: ]{susumuy@mech.kth.se}
\author{Blandine Allais$^2$}
\author{Wouter van der Wijngaart$^3$}
\author{Fredrik Lundell$^1$}
\author{Gustav Amberg$^{1,4}$}
\author{Shervin Bagheri$^1$}

%\author{Blandine Simon}

\affiliation{$^1$Department of Engineering Mechanics, Royal Institute of Technology, SE-100 44 Stockholm, Sweden}%
\affiliation{$^2$\'Ecole Normale Sup\'erieure de Lyon, Lyon, France }%
\affiliation{$^3$ Division of Micro and Nanosystems, Royal Institute of Technology, SE-100 44 Stockholm, Sweden}%
\affiliation{$^4$ Södertörn University, Stockholm, Sweden}%
\date{\today}% It is always \today, today,

\begin{abstract}
The impact of liquid drops on a rigid surface is central in cleaning, cooling and coating processes in both nature and industrial applications. However, it is not clear how details of pores, roughness and texture on the solid surface influence the initial stages of the impact dynamics.
Here, we experimentally study drop impacting at low velocities onto surfaces textured with asymmetric (tilted) ridges.
We define the line-friction capillary number $Ca_f={\mu_f V_0}/{\sigma}$ (where $\mu_f$, $V_0$ and $\sigma$ are the line friction, impact velocity and surface tension, respectively) as a measure of the importance of the topology of surface textures for the dynamics of droplet impact.
We show that when $Ca_f \ll 1$, the contact line speed in the direction against the inclination of the ridges is set by line-friction, whereas in the direction with inclination the contact line is pinned at acute corners of the ridge. 
When $Ca_f \sim 1$, the pinning is only temporary until the liquid-vapor interface reaches to the next ridge where a new contact line is formed. 
Finally, when $Ca_f\gg 1$, 
the geometric details of non-smooth surfaces play little role.       
\end{abstract}

\maketitle

%\tableofcontents

\section{Introduction}

Droplet impact on a solid surface is essential in technological applications such as spray coating and cooling\cite{Dykhuizen1994, Josserand2016}, pesticide deposition\cite{Bergeron2000, Liu2017NRM}, and inkjet printing\cite{AttingerJHT2000, Minemawari2011}. 
The complex fluid-surface interaction during the impact -- which includes splashing\cite{MUNDO1995151, XuPRL2005, josserand_lemoyne_troeger_zaleski_2005, DriscollPRL2011, RibouxPRL2014} and trapping of a thin gas film underneath the droplet\cite{MandrePRL2009,BouwhuisPRL2012, RoelandSM2014, VisserSM2015} -- has been studied theoretically\cite{RoismanPRSLA2002, Attane2007PoF, WildemanJFM2016, EggersZaleskiPoF2010}, numerically\cite{EggersZaleskiPoF2010, SchrollPRL2010, WildemanJFM2016, WANGJNFM2017}, and experimentally\cite{clanet_2004, KANNAN2008694, LaanPRA2014, LeeLangmuir2016, LeeJFM2016, LIN201886}. These studies have established useful scaling laws of maximal deformation, which  among other things, are reviewed in \cite{Yarin2006, Josserand2016}.

The influence of surface roughness and microstructures on drop impact has also been studied extensively focusing on different aspects, such as splashing\cite{KANNAN2008694, Lei2007, RobsonSM2016}, bouncing\cite{TranSM2013, TsaiLangmuir2009, Gauthier2015, BartoloEPL_2006}, trapped gas film under the droplet\cite{RoelandSM2014}, and the maximum spreading radius\cite{LeeLangmuir2016, LeeJFM2016, GuemasSM2012, VaikuntaLangmuir2016}. These studies have reported that surface topology influences the spreading and even small roughness delays spreading at a low impact velocity\cite{LeeLangmuir2016}. 
However, it is not completely understood which  microscopic features of a complex surface texture have the largest influence on droplet impact.
%and its subsequent spreading evolution 
%are not completely understood. 
%In particular, the initial spreading after droplet impact remains to be investigated for complex microstructures. 
%Further investigations are thus needed to gain a better understanding of how textured surfaces control droplet impact. 

One example of a complex surface is an symmetric textured surface, i.e. where the unit structure (post, ridge, rising, etc) is not mirror symmetric with respect to the vertical line passing through the center of structure. 
Asymmetric surface textures are used by natural organisms to control
approaching rain drops\cite{Liu2017NRM}. For example, the slanted microgrooves on the peristrome of the "pitcher plant" \textit{Nepenthes alata}\cite{Chen2016, Bohn2004, Bauer2008},
do not only assist to maintain the surface wetted, but they also prevent drops from falling into the pitcher tank\cite{Yu2018}. Although these asymmetric surface structures have been mimicked for technical applications such as oil-water separation\cite{Li2016} and raindrop shielding\cite{Yu2018}, their influence on droplet impact is not  fully understood.

Here, we perform experiments of a droplet impacting a surface with asymmetric microstructures. We measure the spreading radius in different surface-parallel directions and  
quantify the droplet asymmetry by introducing a line-friction capillary number $Ca_f=\mu_f V_0/\sigma$, where $V_0$ and $\sigma$ are the impact velocity and surface tension, respectively, and $\mu_f$ is the local friction at the moving vapor/liquid/solid phase contact line. As $\mu_f$ constitutes the key ingredient in our analysis (in contrast to earlier models\cite{Attane2007PoF, WildemanJFM2016, LeeJFM2016}), we first briefly summarize the notion of contact-line friction, before  discussing the scope of the present study.

\subsection{Contact-line friction}
When a moving contact line exhibits a dynamic contact angle different from the static value we expect a local dissipation at the contact line. de Gennes 1985\cite{deGennes1985} (eq. 4.71, p860) introduced a local dissipation proportional to $\mu_f U^2$ near the moving contact line, where $U$ is the contact line speed and $\mu_f$ is a 'simple friction coefficient' with the same dimensions as viscosity (denoted $\eta_l$ in de Gennes' original paper). This dissipation is expected from fundamental principles of thermodynamics, and it can have different molecular or hydrodynamic origins. Assuming a microscopic cut-off region where fluid slip is allowed\cite{BocquetFD1999}, the dissipation due to slip and viscous friction in the vicinity of the contact line can be viewed as a local dissipation. Under different circumstances, the moving contact line can be treated as a thermally activated process, which is the basis for the molecular kinetic theory (MKT)\cite{BLAKE1969421, BLAKE20061}. See the recent reviews\cite{Bonn2009, Snoeijer2013} for discussions of these and other possibilities.

Regardless of its molecular origin, the parameter $\mu_f$ can be treated as a macroscopically relevant parameter that characterizes the contribution to the total dissipation from processes that are local to the contact line region. As such, it is expected to depend on the combination of the liquid and the substrate properties, as well as on the local dynamic contact angle, but not otherwise on the flow geometry. Equivalent parameters have been introduced and used in the literature, for instance as a linearization of an assumed smooth dependence of contact line speed on dynamic contact angle\cite{Weiland1981}. 
Yue and Feng discussed contact line dissipation in the Cahn-Hilliard model, and derived the resulting relation between contact line speed and the dynamic contact angle. 
Their relation, in our notation, is\cite{Yue2011}
\begin{equation}
	 \frac{3}{2 \sqrt{2}} \frac{\cos{\theta_e} - \cos{\theta}}{\sin{\theta}} =  \frac{\mu_f U}{\sigma},
	 \label{eq:contactangle_eq}
\end{equation}
where $\theta_e$ is the static contact angle, $\theta$ the dynamic contact angle, and $\sigma$ the surface tension.

The contact line friction coefficient can be measured experimentally\cite{Duvivier2011, Vo2018, HONG2014292} or estimated by parameter fitting of numerical simulations to experiments\cite{CarlsonPRE2012, Lee2019}. Steen\cite{Steen2018,Steen2020} recently used driven droplet oscillations to estimate the magnitude of the contact line friction coefficient.
The values of the line friction parameter in previous studies are in the order of 0.1 Pa$\cdot$s for water and increase in proportion to the square root of the liquid viscosity up to $\sim$ 1 Pa$\cdot$s\cite{Do-Quang2015, CarlsonPRE2012, Vo2018}. 
Since $\mu_f$ is much larger than liquid viscosity for most aqueous solutions\cite{CarlsonPRE2012,Do-Quang2015,Duvivier2011}, the contact line friction plays a particularly dominant role in dynamic and forced wetting applications. 

%\subsection{Line friction in spontaneous spreading}
%In contrast to impacting drops, 
The sensitivity of the line friction parameter to surface properties has been investigated thoroughly within the context of spontaneous spreading (i.e. zero impact speed). 
The relevant non-dimensional number in liquid spreading is the line-friction Ohnesorge number
$Oh_f = \muf/\sqrt{\rho \sigma R_0}$\cite{Do-Quang2015}, where $\rho$ and $R_0$ are density and the initial radius of the droplet, respectively.
The line-friction Ohnesorge number quantifies the contribution of the line friction dissipation to the total kinetic energy. One may therefore expect that when $Oh_f\gg 1$ the contact line speed is strongly influenced by the properties of the substrate and in particular the details of the surface geometry.
In this surface-sensitive regime, Carlson \textit{et al.}\cite{CarlsonPRE2012} have shown that when the time is normalized with the time scale based on the line friction parameter, the initial rapid spreading of different droplets on smooth surfaces nearly collapse into one curve. For non-smooth surfaces, one may define an effective line friction parameter that takes geometric surface details into account. Based on this parameter, one may normalize time such that the spreading curves of different droplets on microstructures exhibit nearly the same scaling\cite{Wang2015, Lee2019, Yada2019}.

\subsection{Scope of the present study}
For droplet impact on smooth surfaces, Wang \textit{et al.}\cite{YuliCarlson2017} rescaled previous experimental data with contact-line friction to demonstrate that line friction limits the maximum spreading radius $\betamax$. They suggested the scaling $\betamax \sim (Re\mu/\muf)^{1/2}$, where $\mu$ is liquid viscosity and $Re$ is Reynolds number. However, to the best of our knowledge, %no study has linked the geometrical features of a particular surface to the spreading resistance imposed by line friction during impact.
no study has linked the spreading mechanism on microstructures to the spreading resistance involving the line friction during droplet impact.

In our previous work\cite{Yada2019}, the spontaneous spreading  of a droplet on slanted microstructures (see inset in Fig.~\ref{fig:setup}a) was explained by mechanisms referred to as slip, stick and leap.
The spreading in the direction against the inclination (indicated by red arrow in Fig.~\ref{fig:setup}a) was driven by the slip mechanism, i.e. a so-called ``capillary spreading'' driven by uncompensated Young's force (Eq.~\ref{eq:contactangle_eq}). In the direction with the inclination (indicated by blue arrow in Fig.~\ref{fig:setup}a), the contact line motion could be explained by a combination of slip, stick and leap; the contact line is pinned at the acute corner of the surface microstructures and the average spreading velocity is set by a combination of the capillary spreading on the flat fraction of the surface and "leaping" of the contact line to the next rise of the surface after the pinning. Here, a length scale separation between the droplet size and the microstructures is assumed so that the spreading mechanisms can be considered local at the contact line.

\begin{figure}[t]
    \centering
    \includegraphics[width=0.6\textwidth]{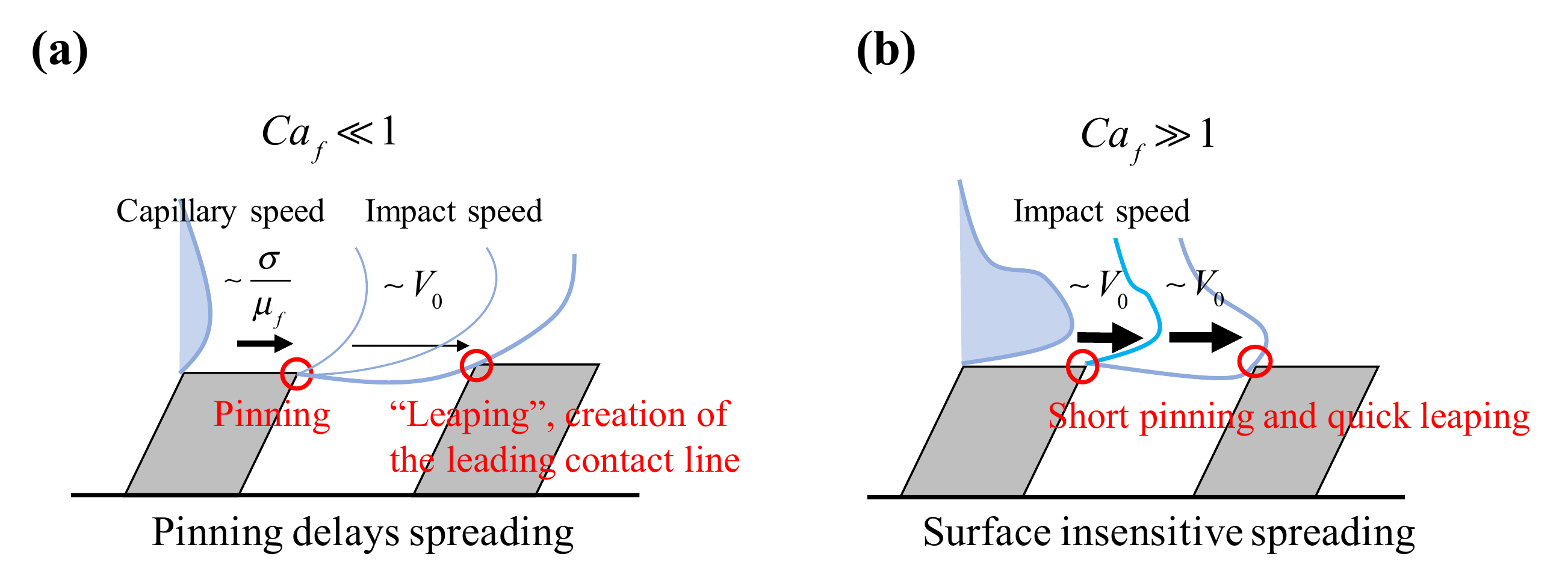}
    \caption{
    Schematics of two limits of droplet spreading on slanted ridges immediately after impact. When $\Camuf \ll 1$, line friction and pinning limit the spreading (a). When $\Camuf \gg 1$, fast leaping between ridges results in a  surface-insensitive spreading (b).
    }
    \label{fig:Pinning}
\end{figure}

In this work, we investigate the same microstructured surface as studied by \cite{Yada2019}, but now for impacting drops, which introduces the impact velocity $V_0$ as an additional parameter.
This allows us to define a new measure of the spreading delay by the surface structures that consists of the ratio between $V_0$ and the characteristic velocity $\sigma/\mu_f$. As shown in Fig.~\ref{fig:Pinning}b, a large  impact speed $V_0$ results in a fast leap of the contact line to the next ridge, which effectively means that the underlying microscropic features of the surface geometry have a small influence on droplet impact.
On the other hand, when $V_0$ is very small compared to $\sigma/\mu_f$ the contact line motion is significantly influenced by both  pinning and line friction  (Fig.~\ref{fig:Pinning}a). 
We therefore propose that the line-friction capillary number, 
$\Camuf = \muf \Vz/\sigma$
is the relevant non-dimensional number to characterize the influence of asymmetric surface geometry on droplet impact. 
%
%to characterize the ratio of the capillary spreading velocity and the characteristic spreading velocity associated with the impact velocity is desirable to quantify the spreading delay by the surface structures 
Note that, despite the fact that the impact of spherical drop on two-dimensional ridges is a three dimensional problem, this study focuses on the local two-dimensional spreading across the asymmetric ridges. This can be motivated by the fact that the curvature of the liquid-vapor interface in the cross-sectional plane is much smaller than the curvature in the horizontal plane. %Therefore,  three-dimensional effects are not significant near the leading contact line crossing over the microstructures and one can focus on the direction of the inclination of the ridges.

%This number is the ratio of impact velocity $\Vz$ to the capillary line friction velocity $\sigma/\muf$, which we call "capillary spreading velocity". 
%When $\Camuf \ll  1$, the spreading in between the surface rise is slower than the spreading on the tip of the microstructure. The leaping takes significantly longer time than the capillary spreading on the flat fraction of the surface, leading to hindered spreading (Fig.~\ref{fig:Pinning}a).
%When $\Camuf \gg 1$, the spreading is completely governed by the impact and insensitive to the surface (Fig.~\ref{fig:Pinning}b). When $Ca_f \sim 1$, the pinning is only temporary as the liquid-vapor interface moves to next ridge where a new contact line is formed. 

\section{Method}
\subsection{Experimental setup}
\begin{figure}[t]
    \centering
    \includegraphics[width=0.45\textwidth]{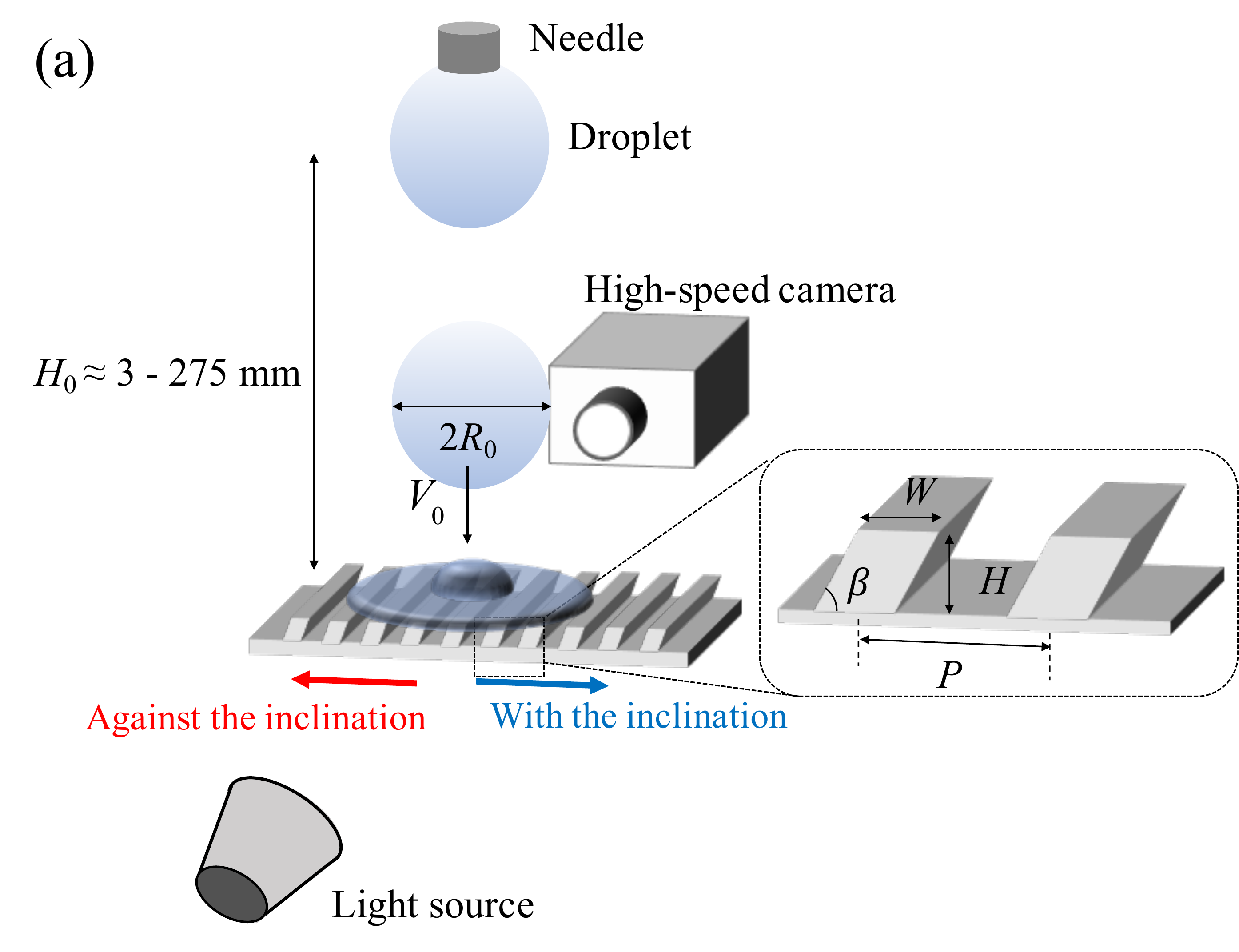}
    \includegraphics[width=0.35\textwidth]{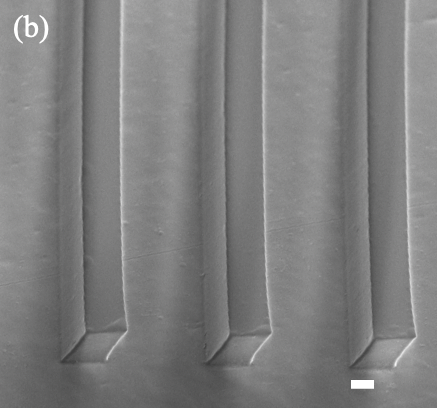}
    \caption{(a) Schematic figure of the droplet impact experiment. (b) Scanning electron microscopy image of the inclined microstructures. The scale bar indicates 10 $\mathrm{\mu}$m}
    \label{fig:setup}
\end{figure}

Impact sequences of liquid droplets are observed with a high-speed camera (Dantek Speedsense M) at a frame rate of 8000 $s^{-1}$ with spatial resolution of 15 $\mu$m. A schematic of the experimental set up is shown in Fig.~\ref{fig:setup}(a). A liquid droplet is formed on the tip of a needle with outer diameter of 0.31 mm (Hamilton, Gauge 30, point style 3) at a height $H_0$ from surfaces to spread on. The liquid is pumped by a syringe pump (Cetoni, neMESYS 1000N) at a small flow rate (0.10 $\mu$l/s). When the growing droplet has reached a certain radius, it pinches off from the needle and is accelerated by gravity and hits the substrate with an impact velocity $\Vz$. The impact velocities, which are varied by changing the distance from the substrate to the needle $H$, are estimated from images before the droplet makes contact with the substrate. The height $H_0$ is varied from 3 mm to 275 mm, which leads to the velocities from 0.16 m/s to 2.3 m/s (table \ref{tab:Height}). 
Spontaneous spreading corresponding to $\Vz$= 0 m/s ($H_0$=0 mm) is also measured.
Fluid properties were varied by mixing de-ionized water, ethanol and glycerol to change viscosity and surface tension. 
We label mixtures of water, glycerol and ethanol (weight ratio of 1:2:1) and water and glycerol (weight ratio of 1:2) as "aq.\ glycerol-ethanol" and "aq.\ glycerol", respectively. 
Fluid properties are shown in Table \ref{tab:property}. 
Viscosity and surface tension are measured with a viscometer (Brookfield) and TD 2 tensionmeter (LAUDA), respectively.

\begin{table*}
    \centering
    \begin{tabular}{lccccccc}
    \hline
         $H_0$(mm) & 3 & 5  & 10  & 25 & 40 & 135 & 275 \\
         \hline
        $V_{0}$(m/s)  & 0.16 & 0.25 & 0.37 & 0.69 & 0.87 & 1.6 & 2.3 \\
        $Ca_f$ for water & 0.27 & 0.42 & 0.62 & 1.2 & 1.5 & 2.7 & 3.8 \\
        $Ca_f$ for aq. glycerol-ethanol & 0.67 & 1.0 & 1.5 & 2.9 & 3.6 & 6.7 & 9.6 \\
        $Ca_f$ for aq. glycerol & 0.89 & 1.4 & 2.1 & 3.8 & 4.8 & 8.9 & 12.8 \\
    \hline
    \end{tabular}
    \caption{List of the heights $H_{0}$, the impact velocities $\Vz$, and the friction capillary number $Ca_f =\muf \Vz/\sigma$.
    }
    \label{tab:Height}
\end{table*}

\begin{table*}
    \centering
    \begin{tabular}{lccccccccc}
    \hline
         Label & \textrho &\textmu & \textsigma & $R_0$ & $\theta_e$ &$\theta_a$ &$\theta_r$ & $\mu_f$ & $\sigma/\mu_{f}$\\
         & (kg/m$^3$) & (mPa $\cdot$ s) & (mN/m) & (mm) & ($^\circ$) & ($^\circ$) & ($^\circ$) & (Pa $\cdot$ s) & (m/s)\\
    \hline 
         Water & 992 & 0.997 & 72 & 1.1 & 50 & 70 & 27 & 0.12 & 0.60 \\ 
        Aq.\ glycerol-ethanol & 1077 & 11.7 & 34 & 0.9 & 34 & 59 & 22 & 0.14 & 0.24\\
        Aq.\ glycerol & 1170 & 15.7 & 63 & 1.0 & 54 & 66 & 28 & 0.36 & 0.18\\
    \end{tabular}
    \caption{Liquid properties, density $\rho$, kinematic viscosity $\mu$, surface tension $\sigma$, initial radius $R_0$, static, advancing, and receding contact angle $\theta_e$, $\theta_a$, $\theta_r$, line friction parameter $\mu_f$, and capillary spreading velocity $\sigma/\mu_{f}$. }
    \label{tab:property}
\end{table*}

\subsection{Surface preparation}
The substrates studied are made from Ostemer 220 (Mercene Labs), a UV-curing Off-Stoichiometry-Thiol-Ene (OSTE) resin\cite{Carlborg2011}. The resin enables to fabricate inclined micropatterns by exposing UV light at an oblique angle. 
The surfaces are prepared in three steps. 
First, a base OSTE layer is prepared on a smooth plastic film.
Second, inclined microridges are patterned on the base OSTE layer by exposing ultraviolet light through a patterned mask. Finally, after cleaning uncured OSTE in an acetone bath, hydrophilic surface modification using 1\% hydroxylated methacrylate (2-Hydroxyethyl methacrylate, Sigma Aldrich) solution in Isopropanol with 0.05\% benzophenone (Sigma-Aldrich) initiator is performed to achieve partial wetting so that the static contact angle on a flat surface is $50^{\circ}$ for de-ionized water. 
Advancing and receding contact angles are measured with the sessile drop method\cite{EralCPS2013}. The sessile droplet with the initial volume $5 \mu l$ is pumped and drained slowly to measure advancing and receding contact angle, respectively. The contact angle right before the contact line starts to advance (recede) is defined as the advancing (receding) contact angle.
The inclination of the ridges $\beta$ is 60$^\circ$. Surface structures are characterized with scanning electron microscopy and the width $W$ is 20 $\mu m$, the pitch $P$ is 60 $\mu m$, and the height $H$ is 20 $\mu m$, as shown in Fig.~\ref{fig:setup}(b).

\subsection{Estimates of line friction parameter}
Experiments of a droplet spreading on a flat OSTE surface are modelled numerically to determine the line friction parameter.
The line friction parameter is determined by fitting the spreading curve with the experiments. 
Spreading of a droplet on a flat surface is experimentally observed with a high-speed camera at a frame rate of 52000 $s^{-1}$ and the spreading radius and the spreading time are recorded. 
To enhance sensitivity to the line friction parameter\cite{Do-Quang2015}, we reduce the initial radius to 0.4 mm. 

Navier-Stokes-Cahn-Hilliard equations are solved using in-house software ``FemLego'' to obtain the spreading radius for different values of $\muf$. 
FemLego is an adaptive finite element toolbox where weak formulation of partial differential equations is defined on a MAPLE worksheet\cite{AMBERG1999257}. 
The numerical model is composed of the Navier-Stokes equations and the Cahn-Hilliard equation;
 \begin{equation}
	\rho(c) \frac{D\textit{\textbf{u}}}{Dt} = - \frac{1}{Re} \nabla p+ \frac{1}{Re} \nabla \cdot \mu(c) (\nabla \textit{\textbf{u}}+\nabla^T \textit{\textbf{u}})-\frac{c \nabla \phi(c)}{Ca_{\mu}\cdot Cn \cdot Re} ,
	\label{eq:NS}
\end{equation}		
\begin{equation}		
	\nabla \cdot \textit{\textbf{u}}=0	,
	\label{eq:continuum}
\end{equation}	
 \begin{equation}
	 \frac{Dc}{Dt} = \frac{1}{Pe} \nabla^{2}  \phi(c).
	 \label{eq:CH}
\end{equation}		
The Navier-Stokes equations are characterized by the capillary number $Ca_{\mu}$=$\mu U /\sigma$, the Reynolds number $Re$=$\rho UL/\mu$, and the Cahn number $Cn$=$\epsilon/L$, where $\rho$ and $\mu$ are the density and viscosity of the liquid phase, $\sigma$ is the surface tension of the liquid-vapor interface, and $\epsilon$ is the diffuse interface width. 
Moreover, $U$ and $L$ are the characteristic velocity and length of the system, respectively.
Here, capillary velocity $\sigma/\mu$ and the initial droplet radius $R_0$ are chosen as the characteristic velocity and length scale.
The variable $c$ is the phase field variable, where $c =1$ represents the liquid phase, and $c  =-1$ the vapor phase.

In the Cahn-Hilliard equation (\ref{eq:CH}), $\phi$ is the chemical potential of the system defined as $\phi=\Psi^{\prime}(c)-Cn \nabla^2 c$.
%, where $c$ is the phase-field variable ($C =1$ represents the liquid phase, and $C  =-1$ the vapor phase). %and is set to 5.7 $\times 10^{-6}$. 
Here, $\Psi(c)= (c+1)^{2}(c-1)^{2}/4$ is the double well function, where the minimum represents the stable phases for vapor ($c =-1$) and liquid ($c =1$). The Peclet number is defined as $Pe=UL/D$ where $D$ is a mass diffusivity. 

The line friction parameter appears as a boundary condition in the Navier-Stokes-Cahn-Hilliard equations in the form\cite{jacqmin_JFM2000, YueFeng_Pof2011}
\begin{equation}
	-\epsilon \mu_{f}  \frac{\partial c}{\partial t} =\epsilon \sigma \nabla c \cdot \textit{\textbf{n}} - \sigma {\rm cos} (\theta _{e})g^{\prime} (c),
	\label{eq:muf}
\end{equation}	
%where $C$ is the phase field variable ($C =1$ represents the liquid phase, and $C  =-1$ the vapor phase). 
where \(\theta_{e}\) is static contact angle. The polynomial $g(c)= 0.5+0.75c-0.25c^3$ rapidly shifts from $0$ (vapor phase $c =-1$) to $1$ (liquid phase $c =1$). The left hand side of Eq.\ref{eq:muf} models the dissipation at the moving contact line. 

The only unknown parameter in the numerical simulation is the line friction parameter.  
We impose the no-slip boundary condition at the wall for the velocity $u$ and Eq.~\ref{eq:muf} as the wall boundary condition for the phase-field variable $c$. Therefore, the local effect at the contact line is effectively modeled in the line friction parameter in this work.
Fluid properties in Table~\ref{tab:property} are used in the simulations to match to the experimental spreading rates. The mass diffusivity in Eq.~\ref{eq:CH} is fixed to 5.7 $\times 10^{-6} ~m^2s^{-1}$ for all of our simulations. The interface width $\epsilon$ is determined following the guidance to maintain the sharp interface limit\cite{yue_feng_JFM2010}. 
The fitted line friction parameters are reported in Table \ref{tab:property}. The simulations are carried out in the axi-symmetric geometry along the center of the droplet. The line friction parameter increases with kinematic viscosity from 0.12 Pa$\cdot$s (water) to 0.36 Pa$\cdot$s (aq.\ glycerol).

\subsection{Numerical simulation of droplet impact}
The droplet impact on the asymmetric microstructure is numerically modeled to investigate how the liquid-vapor interfaces proceed over the microstructures. The simulations reveal the spreading mechanisms over the microstructures, which can not be observed in the experiments due to lack of spatial and time resolutions.

In cylindrical coordinates, Eqs.~(\ref{eq:NS}-\ref{eq:CH}) with Eq.~\ref{eq:muf} are solved with the properties of a water droplet on Table \ref{tab:property} including the fitted line friction parameter. To reduce the computational cost, the initial radius of the droplet is reduced to $0.3$ mm, while the dimension of the surface geometry is identical to the experiment. The droplet is initialized at the distance of $0.033R_0$ from the solid wall with the initial vertical velocity of 0.8 m/s.  
Video animations are available as Supplemental material.
\begin{figure}[t]
    \centering
    \includegraphics[width=0.85\textwidth]{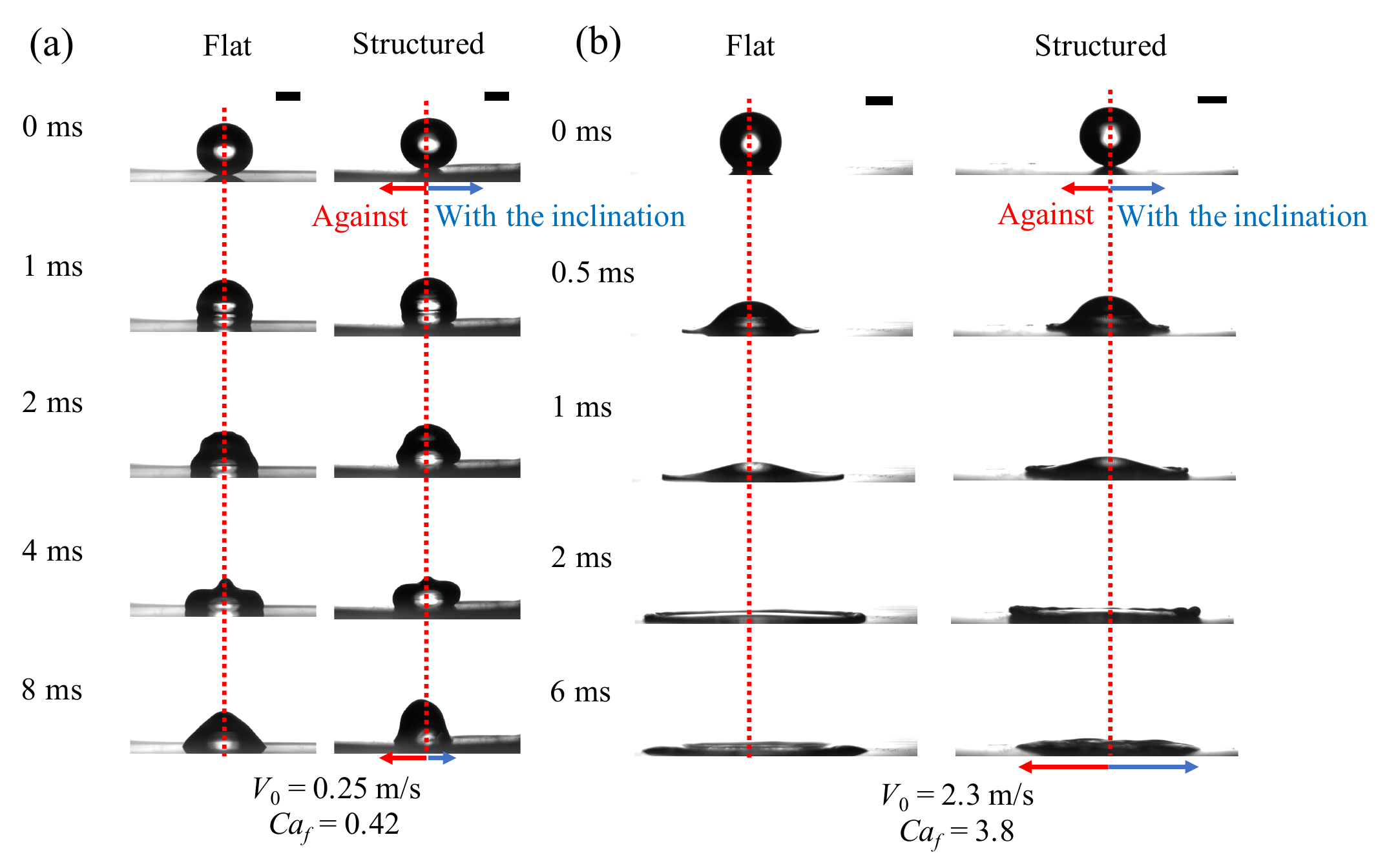}
    \caption{Selected captured images for (a) $\Vz$=0.25 m/s (b) $\Vz$=2.3 m/s of a water droplet. Scale bars represent 1 mm.}
    \label{fig:graphics}
\end{figure}

\section{Results}
\subsection{Comparison between flat and microstructured surfaces}

Figure~\ref{fig:graphics}(a) shows a series of images of a water droplet spreading after impact on the flat surfaces and asymmetrically microstructured surfaces with $\Vz$=0.25 m/s ($\Camuf = 0.42$). We observe that the droplet spreads not only slower on the asymmetric structures compared to the flat surface but also asymmetrically (Fig.~\ref{fig:graphics}a). Specifically, 
the spreading is faster in the direction against the inclination of the ridge than in the direction with the inclination. 

\begin{figure}[tb]
    \centering
     \includegraphics[width=0.8\textwidth]{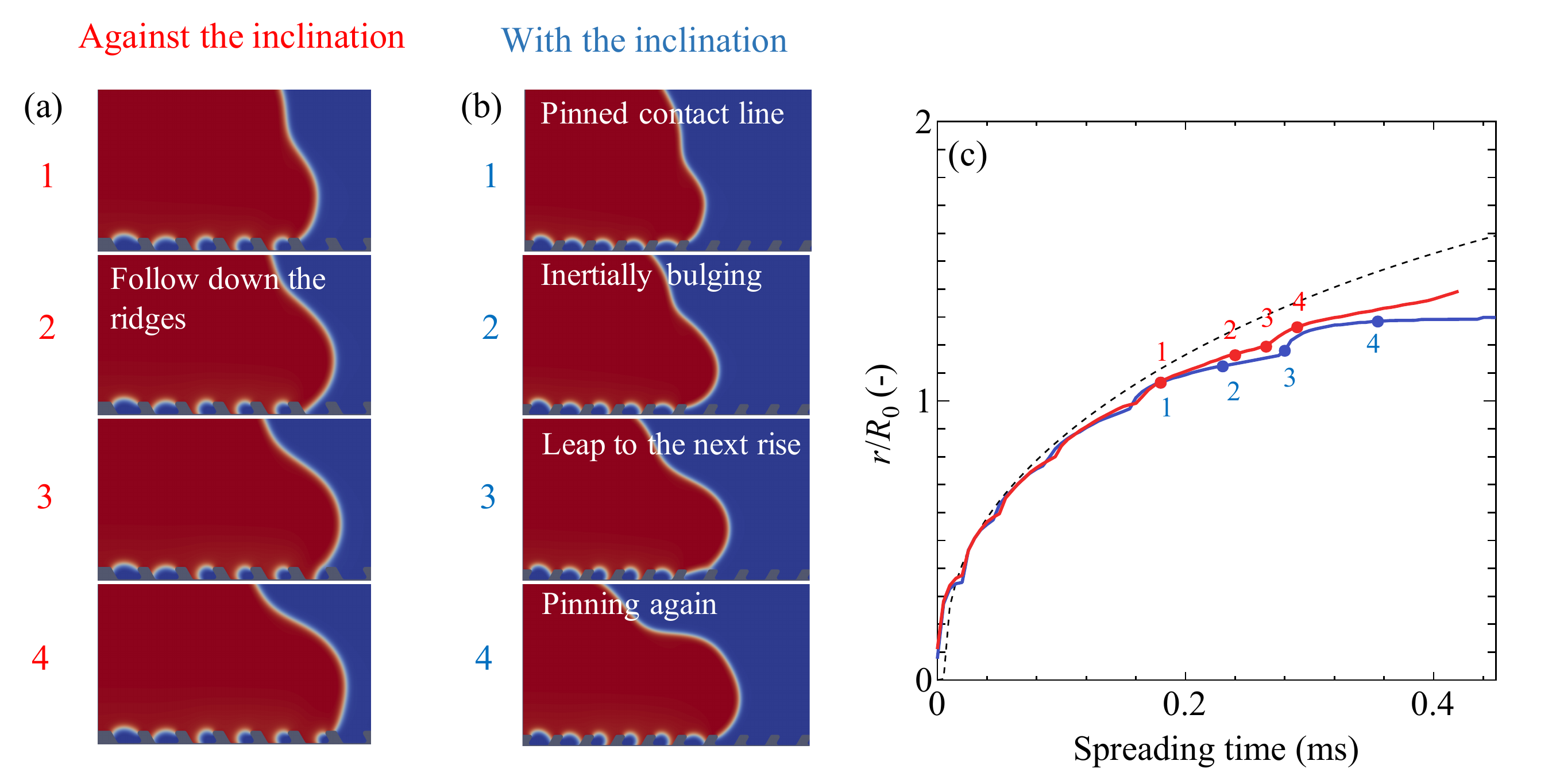}
    \caption{Spreading mechanisms after impact on the inclined microstructures. (a) In the direction against the inclination, the contact line follows the surface structures. (b) In the direction with the inclination, the contact line is pinned at the acute corner of the surface(1, 4). Eventually the liquid-vapor interface reaches to the next rise of the surface (2,3). (c) The simulated spreading radius with respect to time on flat surface (black dash line) and in the direction against (red line) and with (blue line) the inclination on the asymmetric microstructure. The red and blue points 1-4 correspond to the images in (a) and (b), respectively. }
    \label{fig:mechanism}
\end{figure}

%Here, the spreading asymmetry is discussed with the spreading mechanisms revealed with the numerical simulations in Fig.\ref{fig:mechanism}.
Here, the numerical simulations shown in Fig.~\ref{fig:mechanism} reveal the spreading mechanisms on the asymmetric microstructure.
In the direction against the inclination, the contact line follows along the microstructure without pinning. As a consequence, it travels a longer path compared to its flat counterpart and therefore the apparent spreading rate is slightly slower (Fig.~\ref{fig:mechanism}a). 
In the direction with the inclination, the contact line spreads only on the tip of the surface ridges before it is
temporarily pinned at the acute corner of the surface (1, 4 Fig.~\ref{fig:mechanism}b). The  contact angle does not reach the advancing contact angle to move down along the inclined wall and the contact line is suspended.
During pinning, the liquid-air interface is stretched until it reaches the next rise of the surface (2-3, Fig.~\ref{fig:mechanism}b). The spreading in this direction is delayed by the surface geometry compared to the flat surface if the duration of the pinning is longer than the time it would take for the interface to spread over a flat surface. Note that this mechanism is very similar to the slipping mechanism of a droplet on superhydrophobic surfaces observed experimentally with laser scanning confocal microscopy\cite{HansPRL2016}.
At $\Camuf=1.3$, a slight spreading asymmetry is observed in the simulation as shown in Fig.~\ref{fig:mechanism}(c).  
We also note that the simulation is carried out in an axisymmetric geometry. In the experiments, the cavity between the ridges might be filled up with the liquid phase immediately due to three-dimensional effects. The numerical model therefore only provides a qualitative picture of the asymmetric spreading.

Figure~\ref{fig:graphics}(b) shows snapshots of a droplet with $V_{0}$=2.3 m/s ($\Camuf =3.8$) on flat and asymmetric surfaces.
We observe symmetric spreading on the microstructured surface, indicating a small effect of microstructure geometry on liquid spreading.
%Since the impact velocity is much higher than the spontaneous spreading velocity scale, 
In this case, the impact velocity reduces the pinning time and favours the leaping mechanism (Fig.~\ref{fig:Pinning}b).

\begin{figure*}[t]
    \centering
    \includegraphics[width=0.9\textwidth]{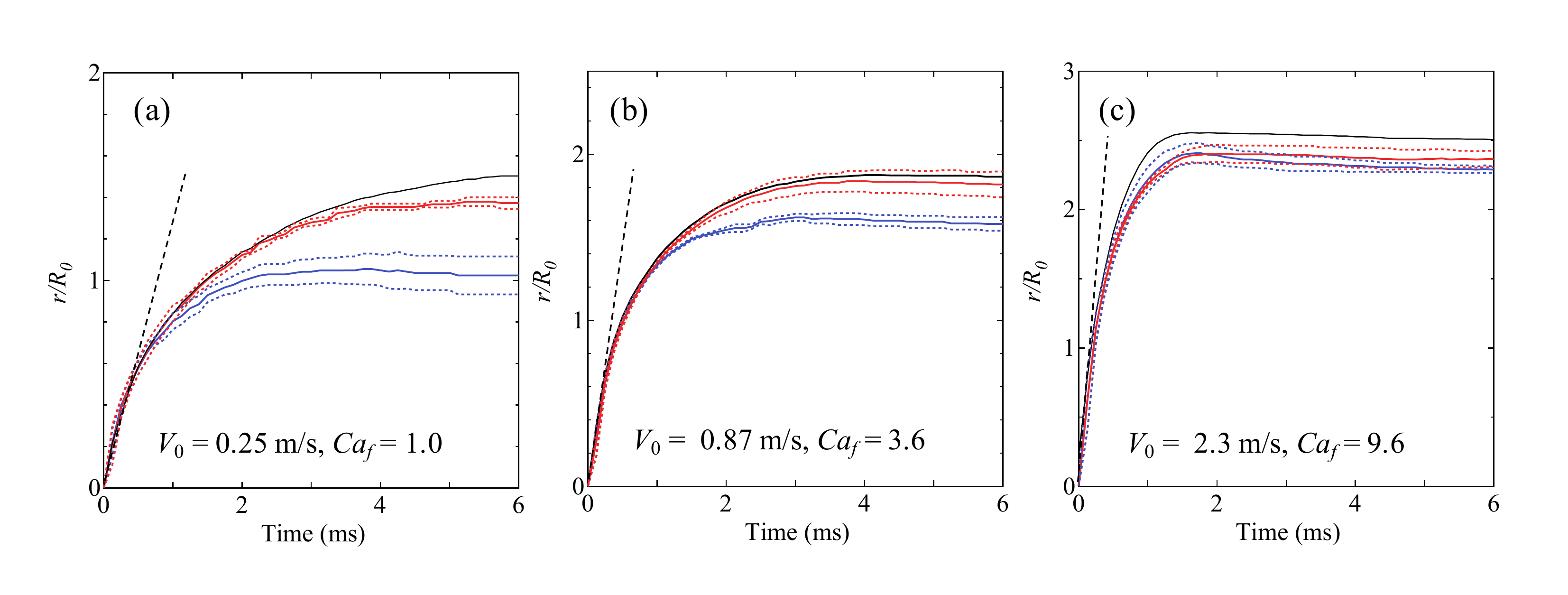}
    \caption{Normalized spreading radius $r/R_0$ of a droplet of aq.\ glycerol-ethanol as a function of time for $V_0 $(a) 0.25 m/s (b) 0.87 m/s, and (c) 2.3 m/s. Black, red, blue curves represent flat surface, the direction against the inclination, and the direction with the inclination, respectively. Dash lines show the initial slope of the spreading curves. The curves are averages of at least 5 repeated measurements. Dotted lines represent the standard deviations.
    }
    \label{fig:Curves}
\end{figure*}

Figure \ref{fig:Curves} shows the spreading curves of droplets immediately after impact of aq.\ glycerol-ethanol with three different impact velocities.
In all three cases in Fig.~\ref{fig:Curves}(a--c), the spreading curves on the flat surface and asymmetric microstructures collapse in the initial phase, until around 1 ms. 
The spreading velocity in this phase -- estimated from the slope of the spreading curve in the initial phase -- is significantly higher than the impact velocity. For example, in Fig.~\ref{fig:Curves}(a) it is $\sim$ 1 m/s, which is a factor of 4 faster than the impact velocity. %and even higher in Fig.~\ref{fig:Curves}(b) and (c). 
The spreading in this very initial phase is fully inertial and essentially independent of the contact line friction and consequently also insensitive to the surface structures.
After the initial phase, the spreading curves in the direction against and with the inclination begin to deviate from each other (Figs.~\ref{fig:Curves}a and b). 
Specifically, the spreading in the direction against the inclination (red curves) closely follows the one of the flat surface (black curves). In this direction, the small reduction in spreading velocity can be attributed to the increase of wetted area of the microsctructured surface compared to the flat surface and not to different spreading mechanisms.  
On the other hand, the spreading in the direction with the inclination (blue curve) is slowed down significantly. At these low impact velocities, this can be attributed to the pinning of the contact line at the acute corner of the structures.

In contrast, in Fig.~\ref{fig:Curves}(c), 
%Moreover, in Figs.~\ref{fig:Curves}(a) and (b), the spreading in the direction against the inclination (red curves) closely follows the one of the flat surface (black curves). In this direction, the small reduction in spreading velocity can be attributed to the increase of wetted area of the microsctructured surface compared to the flat surface and not to different spreading mechanisms.  
for a high impact velocity, the spreading curve in the direction with the inclination approaches the curve of the flat surface. Here, the pinning time becomes shorter and the delay by the surface structure in the direction with the inclination diminishes, as could be expected by $\Camuf \gg 1$. Consequently, the spreading is nearly symmetric on the asymmetric microstructure over the entire spreading and close to the spreading on the flat surface.

\subsection{Maximum spreading radius}
\begin{figure*}[t]
    \centering
    \includegraphics[width=0.85\textwidth]{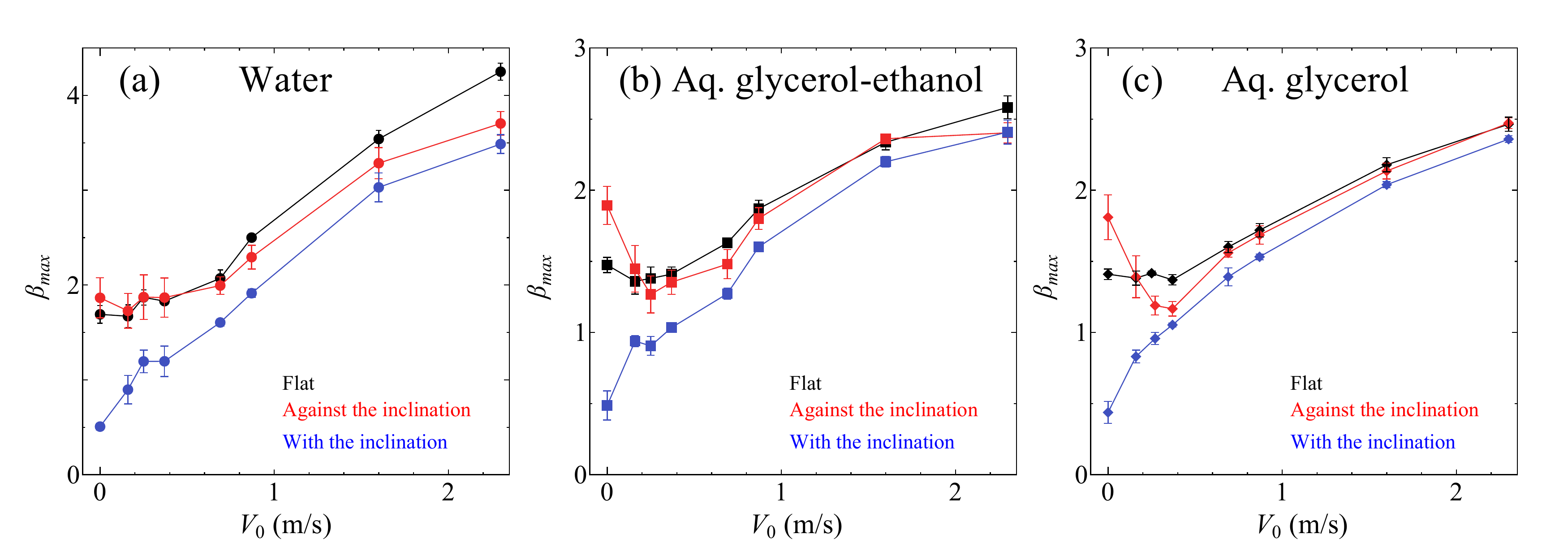}
    \includegraphics[width=0.4\textwidth]{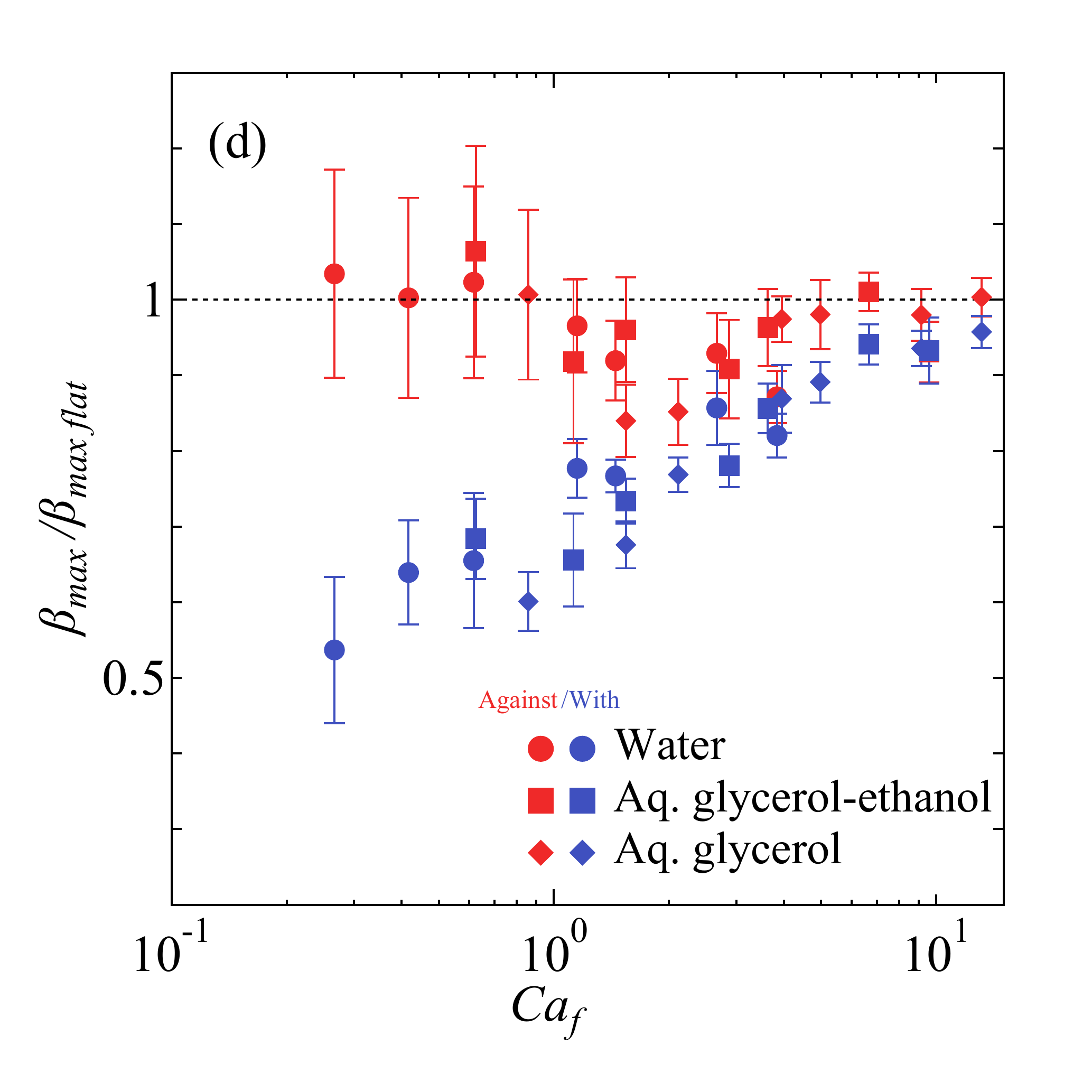}
     \includegraphics[width=0.4\textwidth]{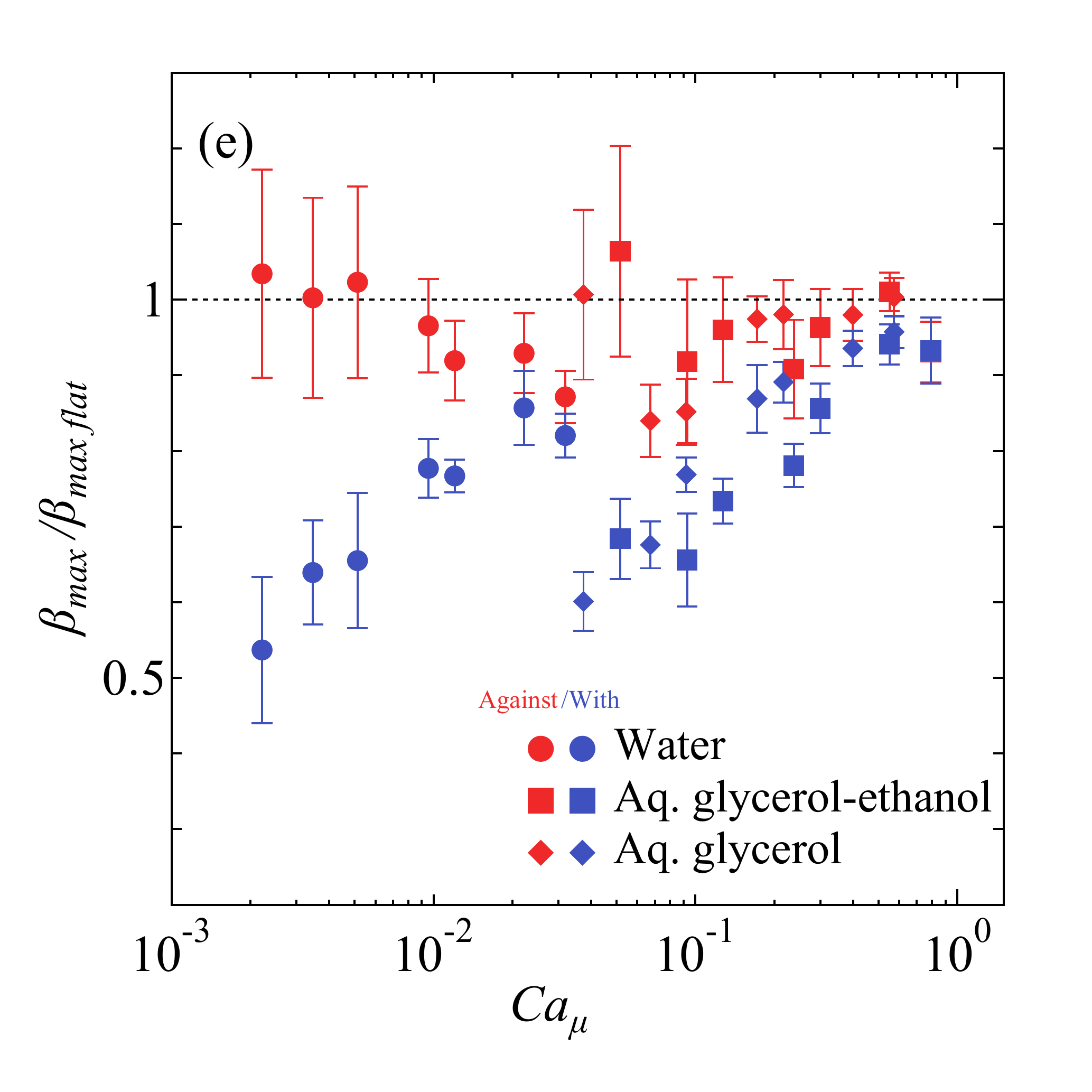}
    \caption{
    (a,b,c) Normalized maximum spreading radius with respect to $\Vz$ of (a) water (b) aq.\ glycerol-ethanol (c) aq.\ glycerol. Black, red, blue marks represent flat surfaces, the direction against the inclination, and with the inclination on the asymmetric microstructures, respectively.
    (d, e) The relative spreading factor to the flat surface with respect to (d) $\Camuf$ and (e) $Ca_{\mu}$. The spontaneous cases are eliminated.
    Error bars in (a-e) indicate standard deviations.
    }
    \label{fig:Rmax1}
\end{figure*}

Figures \ref{fig:Rmax1}(a--c) show the normalized maximum spreading radius, so called ``spreading factor'', $\betamax=R_{max}/R_0$, with respect to the impact velocity. At low impact velocity, the maximum spreading on flat surfaces (black curves) is relatively independent from the impact velocity. This implies that the impact sequence is mainly driven by Young's force (capillary and line friction) similar to the spontaneous spreading of a deposited droplet ($\Vz$ = 0).
The spreading factor increases with impact velocity above $V_0\sim 1$ m/s, as the spreading gradually becomes more dominated by the impact.
On asymmetric microstructured surfaces, the spreading factor in the direction against the inclination (red curve) follows the spreading factor on the flat surface, except for the water droplet with high impact velocity (Fig.~\ref{fig:Rmax1}a).
In contrast, the spreading factor in the direction with the inclination (blue curve) is smaller than the flat surface, but it approaches that of the flat surface as the impact velocity increases. The reduced pinning time with the increased impact velocity is responsible for this trend.

Figure \ref{fig:Rmax1}(d) shows the spreading factor on the asymmetric microstructured surface normalized by the spreading factor on the flat surface with the same impact velocity. The horizontal axis shows the line-friction capillary number. 
The normalized spreading factor in the direction against the inclination is almost constant around 1. Meanwhile, the normalized spreading factor in the direction with the inclination monotonically increases from 0.5 to 1 with increasing $\Camuf$. As a result, the asymmetry in the spreading factor decreases monotonically with increasing $\Camuf$, while for $\Vz$ = 0 m/s, the spreading factor in the direction against the inclination is a factor of 4 larger than in the direction with the inclination (see Fig.~\ref{fig:Rmax1}a--c). It is important to note that the conventional capillary number $Ca_{\mu}=\mu \Vz/\sigma$ does not give monotonic trend (Fig.~\ref{fig:Rmax1}e). As the line friction parameter increases in proportion to the square root of liquid viscosity, more viscous fluid is likely to have higher $\Camuf$ for the same impact velocity. 
%Therefore, it is likely that the spreading of very viscous fluids is insensitive to the asymmetric microstructures discussed in this study. 

\begin{figure*}[t]
    \centering
    \includegraphics[width=0.37\textwidth]{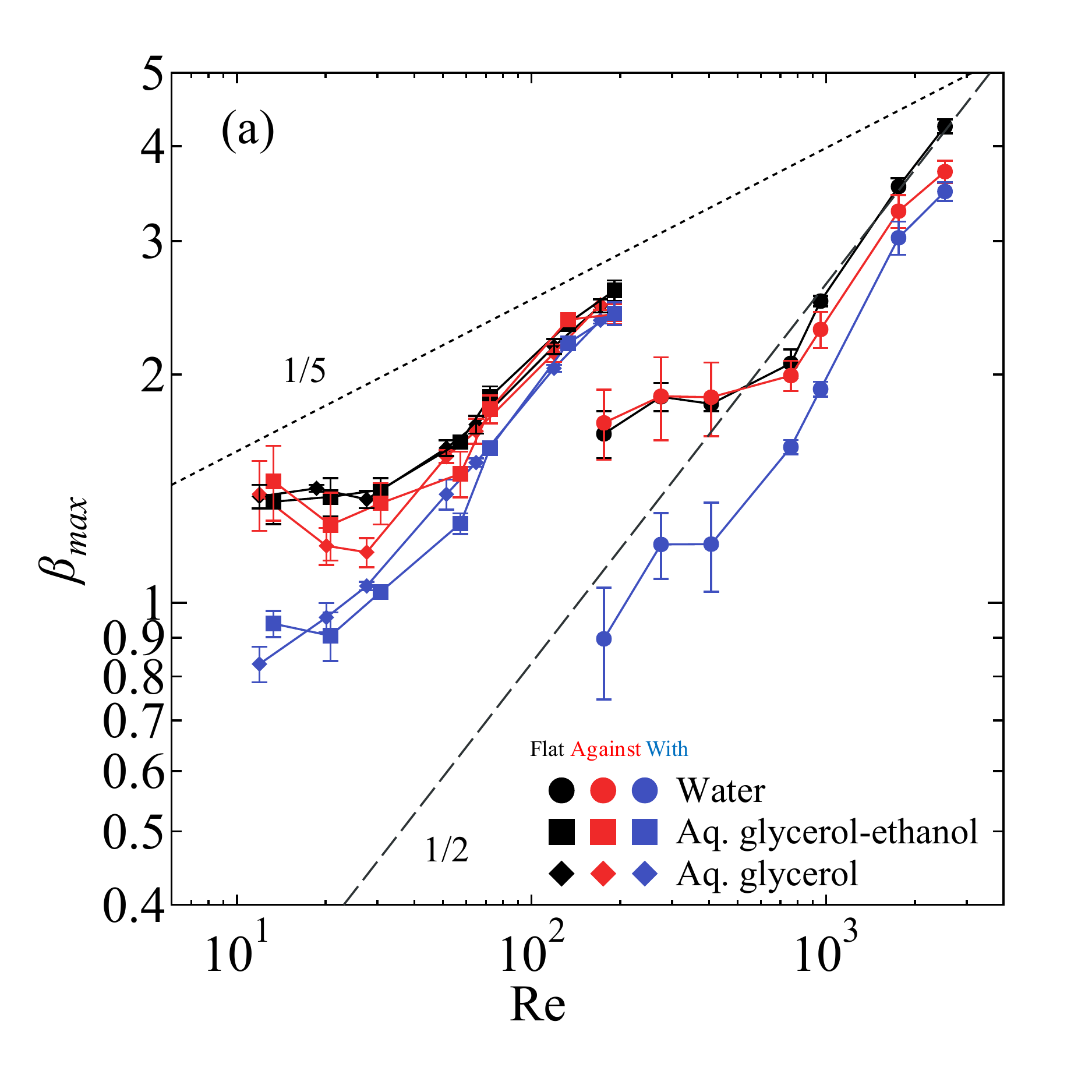}
    \includegraphics[width=0.37\textwidth]{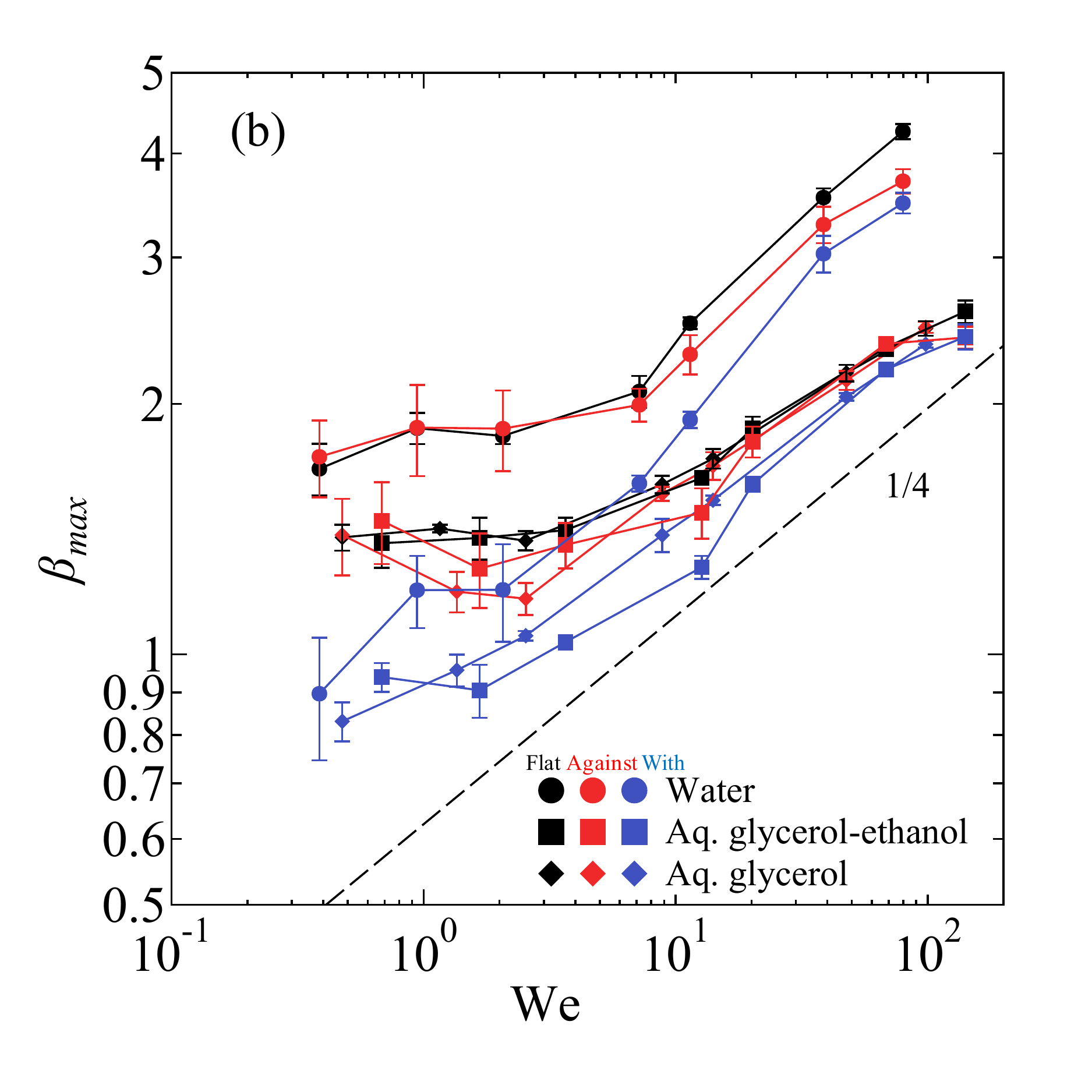}
    \includegraphics[width=0.37\textwidth]{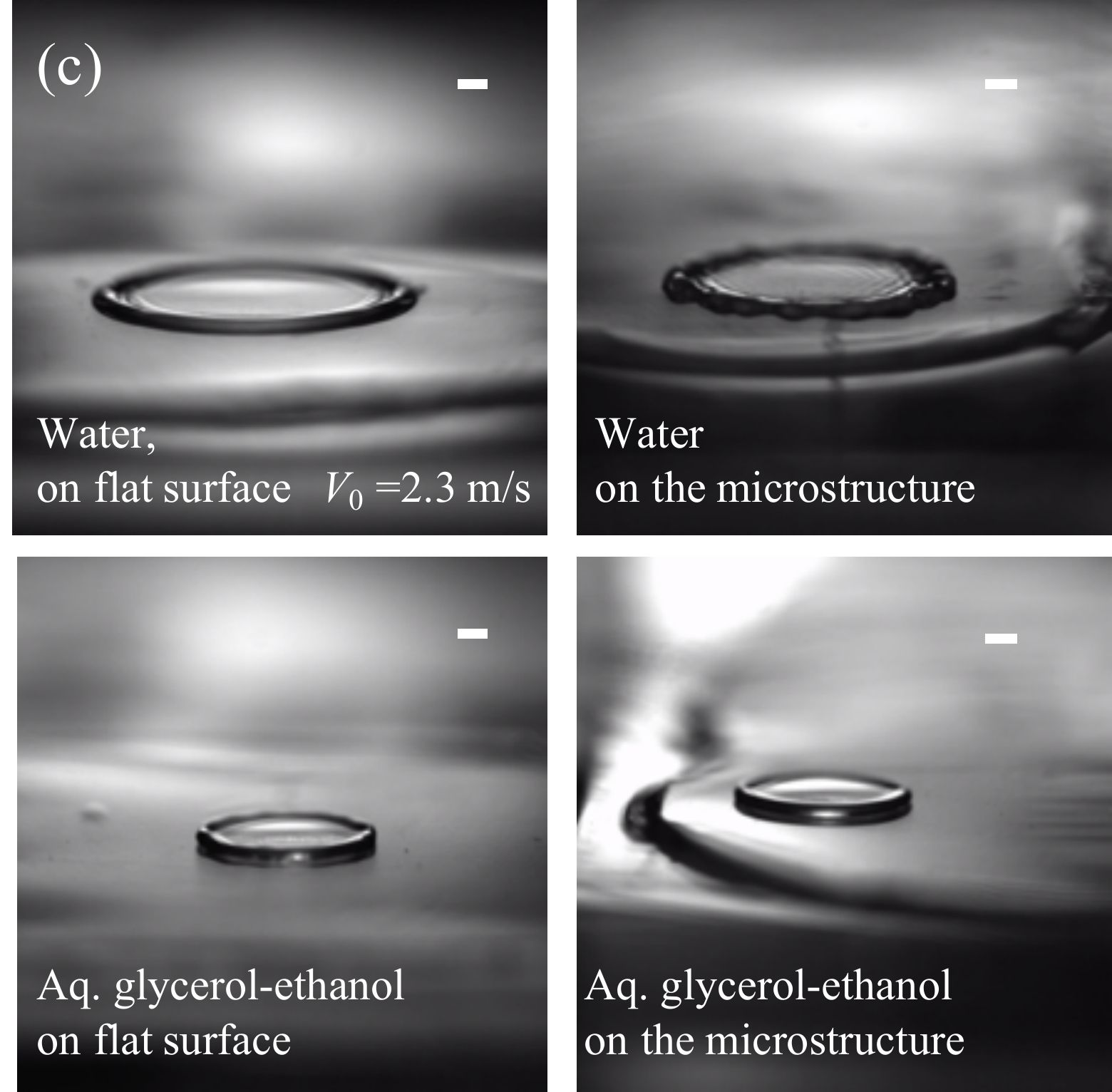}
    \caption{
    (a, b) Normalized maximum spreading radius with respect to (a) Reynolds number $Re = {\rho R_0 V_0}/{\mu}$ (b) Weber number $We ={\rho R_0 V_{0}^2}/{\sigma}$.
    Error bars in (a) and (b) indicate standard deviations.
    (c) Liquid lamella of water (top) and aq.\ glycerol-ethanol (bottom) on the flat surface (left) and the microstructured surfaces (right) at the moment of the maximum spreading radius with $\Vz$ = 2.3 m/s. 
    The images are taken with an oblique angle. Scale bars indicate 1 mm.
    }
    \label{fig:Rmax2}
\end{figure*}

Figures \ref{fig:Rmax2}(a) and (b) show the spreading factor with respect to Reynolds number $Re=\rho \Vz R_0/\mu$ and Weber number $We = \rho R_0 V_{0}^2/\sigma$. 
The well-known relation with the Reynolds number
\begin{equation}
\betamax \sim Re^{\frac{1}{5}}, 
\label{eq:RmaxRe}
\end{equation}
is theoretically derived assuming that the kinetic energy is solely dissipated by viscous dissipation\cite{EggersZaleskiPoF2010, LIN201886, clanet_2004}. Note that this is valid only in the viscous regime (i.e., for low Reynolds number).
 However, the spreading factors in our study do not follow Eq.~\ref{eq:RmaxRe}, but $\betamax \sim Re^{\frac{1}{2}}$, as seen in Fig.~\ref{fig:Rmax2}(a). This also agrees well with previous experimental observations of Lin \textit{et al.}\cite{LIN201886} for high Reynolds numbers. 
Using an energy balance analysis, Wang \textit{et al.}\cite{YuliCarlson2017} proposed the following scaling of the spreading factor,
\begin{equation}
Re \approx \betamax^2 \left ( \frac{\muf}{\mu}+\frac{1}{Ca_{\mu}}+\betamax^3 \right ),
\label{eq:Energy}
\end{equation}
where $Ca_{\mu}=\mu \Vz/\sigma$ is the capillary number based on bulk viscosity. 
The three terms represent the contributions to the energy budget by the contact line dissipation, work done by surface tension, and viscous dissipation, respectively.
Here, the first term is the leading term in Eq.~\ref{eq:Energy} in our study, i.e., $\muf/\mu \gg 1/Ca_{\mu}+\betamax^3$, and we obtain
\begin{equation}
\betamax \sim (Re \mu/\muf)^{\frac{1}{2}}.
\label{eq:RmaxRemuf}
\end{equation}
Note that the exponent in Eq.~\ref{eq:RmaxRemuf} agrees with our experiments. For more viscous fluids, when $(\betamax^3 \gg {\muf}/{\mu} + {1}/{Ca_{\mu}}$, the classical scaling law for the viscous regime (Eq.~\ref{eq:RmaxRe}) is recovered.
%Note that the spreading factor for small $\Vz$ does not follow Eq.~\ref{eq:RmaxRe}. 

Meanwhile, the well-known relation between the spreading factor and Weber number is\cite{clanet_2004, LIN201886}
\begin{equation}
\betamax \sim We^{\frac{1}{4}},     
\label{eq:RmaxWe}
\end{equation}
in the capillary regime with low Ohnesorge number, while a lower exponent (1/6) is reported for viscous fluids with \textit{Oh}=0.585\cite{Attane2007PoF, clanet_2004}. An analysis based on the momentum and mass conservation leads to Eq.~\ref{eq:RmaxWe}\cite{clanet_2004}.
The spreading factor in this work follows Eq.~\ref{eq:RmaxWe} well (Fig.~\ref{fig:Rmax2}b). This is in reasonable agreement with previous studies\cite{clanet_2004, LIN201886} since \textit{Oh} in this study ranges from $3.6 \times 10^{-3}$ to $6.4 \times 10^{-2}$, which is regarded as the capillary regime.
To conclude the scaling analysis, our experimental parameter space is in the capillary regime and the classical scaling law with Weber number is observed. %This is characterized by low Ohnesorge number. 
The classical scaling with the Reynolds number (Eq.\ref{eq:RmaxRe}) must be reconsidered in the capillary regime and the reasonable scaling (Eq.\ref{eq:RmaxRemuf}) is theoretically obtained by applying the energy balance analysis by Wang \textit{et al.}

The spreading factor on the microstructures for high Re (water, $\Vz=2.3$ m/s) does not reach the value on the flat surfaces since the liquid lamella begins to break earlier on the microstructured surfaces compared to flat surfaces. As shown in Fig.~\ref{fig:Rmax2}(c), the water lamella breaks only on the microstructures but not on the flat surface. On the other hand, the lamella of aq.\ glycerol-ethanol is stable at $\Vz$= 2.3 m/s both on the flat and the microstructured surfaces.
A criterion for splash is $K = We \sqrt{Re} \gtrsim 3000$\cite{Josserand2016, MUNDO1995151}. For the water droplet with $\Vz=2.3$ m/s we obtain $K \sim 4000$ which is slightly higher than the critical $K \gtrsim 3000$, while $K \sim 2000$ for the aq.\ glycerol-ethanol. Therefore, the instability of the water lamella in Fig.~\ref{fig:Rmax2}(c) can be understood as onset of a splash induced by the surface structure. It is responsible for the smaller spreading factor on the microstructured surface compared to the flat surface of a water droplet for high impact velocity.

In practical situations such as raindrops\cite{GossardJAOC1992} and inkjet printing\cite{SenJMM2007,NegroSR2018}, the impact velocity can be beyond the velocity we investigate, as high as 10 m/s for raindrops, for example.
In such situations, $\Camuf \gg 1$ is expected and the spreading is insensitive to the organized microstructures.
This implies that the microstructures are not very effective to harness such highly inertial droplets.

\section{Conclusions}
Spreading of a droplet after impact on asymmetrically microstructued surfaces has been experimentally investigated. 
Considering the microscopic spreading mechanisms, the line-friction capillary number $\Camuf = {\muf \Vz}/{\sigma} $ is proposed to
distinguish between symmetric and asymmetric droplet spreading after impact. 
This non-dimensional number describes the ratio
between capillary speed to the impact velocity. For the tilted microscale ridges considered here, the spreading in the direction against the inclination is not very sensitive to the surface structures, while the spreading in the direction with the inclination scales well with $\Camuf$. Consequently, the asymmetry in the maximum spreading radius fades out with increasing $\Camuf$. 
The scaling law for the spreading factor with Weber number ($\betamax \sim We^{{1}/{4}}$) is confirmed to hold for spreading on asymmetric surfaces. However, the scaling law with Reynolds number shows larger exponent than in the classical theories ($\betamax \sim Re^{{1}/{5}}$) The spreading factor in our experiments follows the scaling proposed by Wang \textit{et al.}\cite{YuliCarlson2017}, which takes the energy dissipation at the contact line into account in the energy balance analysis. 
 Further work considering other surface geometries are needed to see if  $\Camuf\lesssim 1$ can be used as a general condition to distinguish between symmetric and asymmetric spreading after droplet impact. 

\section{Acknowledgements}
This work was supported by the Swedish Research Council (VR 2015-04019) %. S.B acknowledges the support of
and by the Swedish Foundation of Strategic research (SSF-FFL6). 
We would like to thank M. Do-Quang for his technical support. 

\bibliography{apssamp}% Produces the bibliography via BibTeX.

\end{document}